\documentclass[%
reprint,
superscriptaddress,
 amsmath,amssymb,
 aps,
]{revtex4-2}
\usepackage{xcolor}
\usepackage{hyperref}

\usepackage{graphicx}
\usepackage{dcolumn}
\usepackage{bm}
\usepackage{array}

\begin{document}


\title{Angle-of-Arrival Determination of Radio-Frequency Fields \\ Using Stark-Shifted Rydberg-EIT Spectra}

\author{Rajavardhan~Talashila}
\affiliation{National Institute of Standards and Technology, Boulder, Colorado 80305, USA}
\affiliation{Department of Electrical, Computer, and Energy Engineering, University of Colorado, Boulder, Colorado 80309, USA}
\author{Zoya Popovic}
\affiliation{Department of Electrical, Computer, and Energy Engineering, University of Colorado, Boulder, Colorado 80309, USA}
\author{Nikunjkumar~Prajapati} 
\affiliation{National Institute of Standards and Technology, Boulder, Colorado 80305, USA}
\author{Noah~Schlossberger}
\affiliation{National Institute of Standards and Technology, Boulder, Colorado 80305, USA}
\author{Christopher~L.~Holloway}
\affiliation{National Institute of Standards and Technology, Boulder, Colorado 80305, USA}
\email[]{christopher.holloway@nist.gov}


\date{\today}

\begin{abstract}
We demonstrate a method for determining the magnitude of the angle of arrival (AoA) of a linearly polarized radio-frequency (RF) field using the angle-dependent spectral amplitudes of AC Stark-shifted electromagnetically induced transparency (EIT) resonances. An off-resonant RF field produces distinct Stark shifts of the $|m_J|$ sublevels, while their relative excitation strengths depend strongly on the relative orientation of the RF and optical polarizations. Under the controlled polarization geometry considered here, this dependence  enables determination of the AoA magnitude without requiring RF phase measurements or spatially separated sensing locations. We experimentally demonstrate the method at RF frequencies of 1.27 GHz, 2 GHz, 3 GHz, and 4 GHz using a subwavelength cesium vapor cell. Within the $20^\circ-40^\circ$ region of highest angular sensitivity, the estimated repeatability-based $1\sigma$ AoA uncertainty ranges from approximately $0.6^\circ$ to $2^\circ$ across the investigated RF frequencies. A model combining Floquet-Shirley calculations of the RF-dressed Rydberg states with a coherent hyperfine-to-fine-structure optical excitation model qualitatively reproduces the experimental observations. These results establish a path toward compact, phase-independent AoA sensing using spatially localized Rydberg-atom spectroscopy.
\end{abstract}

\maketitle


\section{Introduction}

Determining the direction of an incoming radio-frequency time-harmonic plane wave is usually done using the phase and amplitude differences across an antenna array \cite{stutzman_antenna_2013}. As the signal arrives at different times at spatially separated antennas, the true time delay between the arrivals results in a phase difference at the frequency of interest, which in turn can determine the AoA \cite{mailloux_phased_2018}.  Monopulse techniques use the out-of-phase superposition of a received wave on two or more antennas \cite{sherman_monopulse_2011,jenkins_small-aperture_1991}. Usually, more than two antennas are needed to determine AoA around a certain carrier frequency, implying an electrically large receiver for high-resolution AoA. For a broadband AoA system, multiple antenna arrays are needed, further increasing the size \cite{balanis_antenna_2016}.

\begin{figure}
    \centering
    \includegraphics[width=1.0\linewidth]{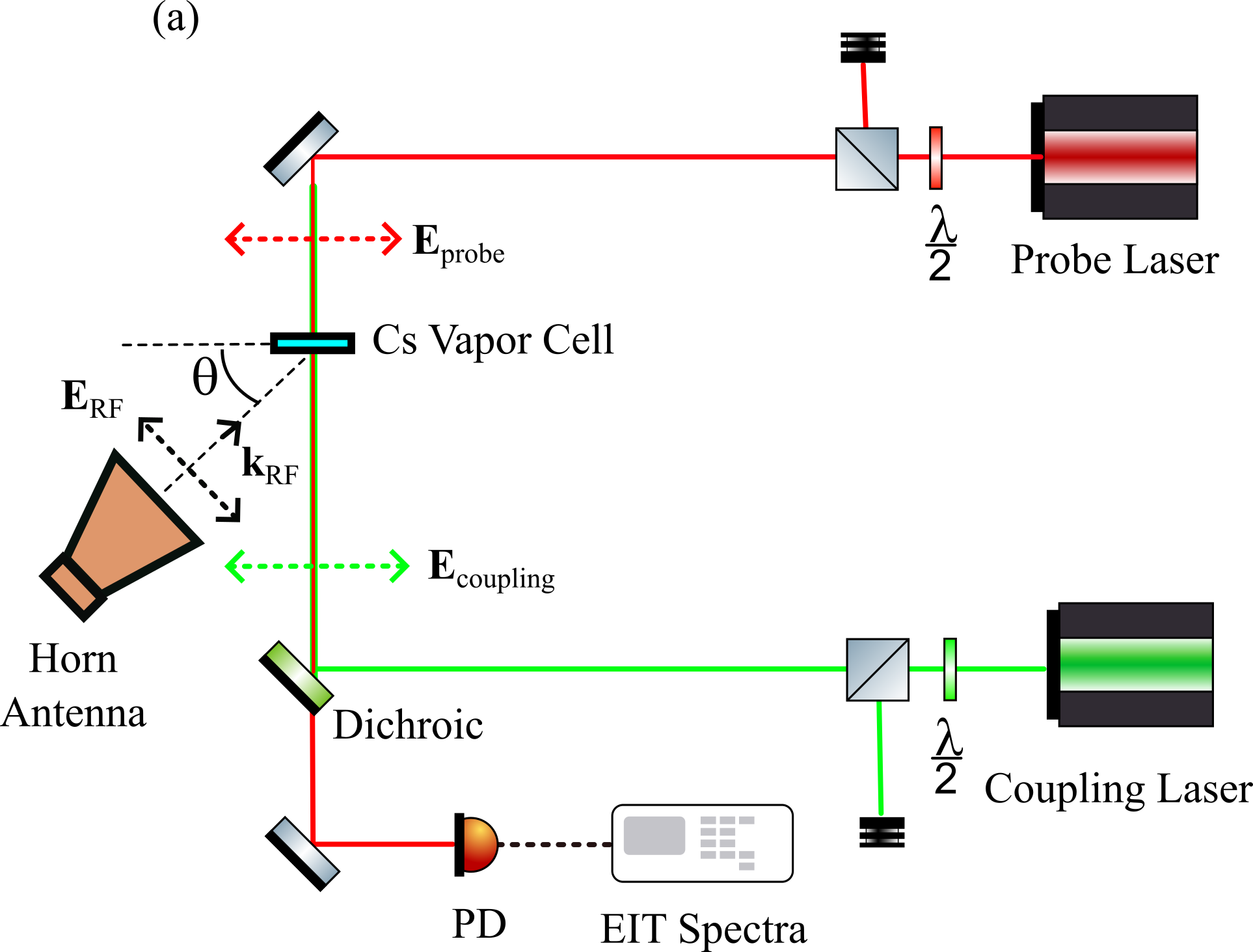}
    \includegraphics[width=0.6\linewidth]{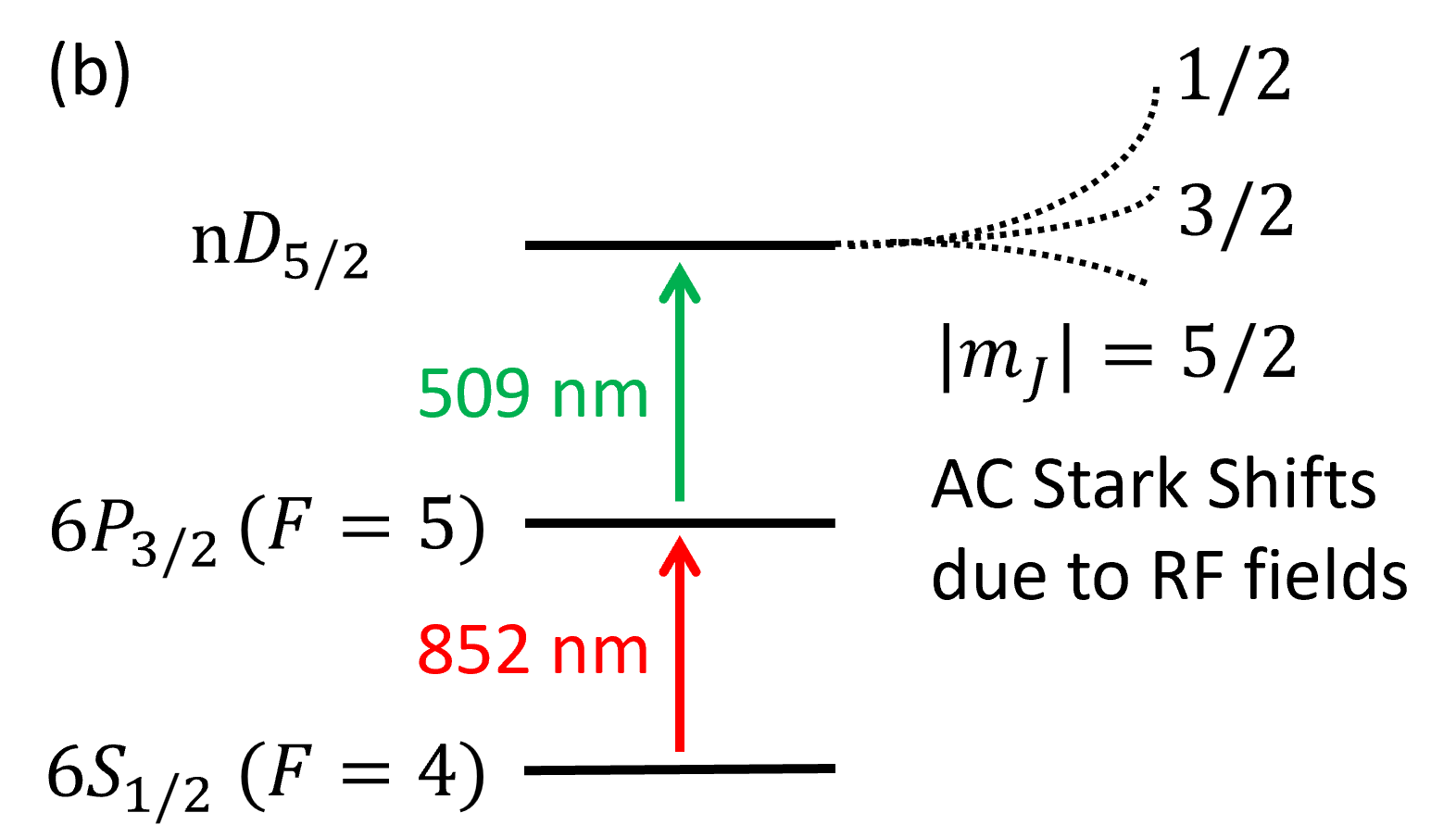}
    \caption{(a) Schematic of the experimental setup. The  RF electric-field polarization  $\mathrm{\mathbf{E}_{RF}}$, and the co-polarized  probe and coupling fields, $\mathrm{\mathbf{E}_{probe}}$  and  $\mathrm{\mathbf{E}_{coupling}}$, lie in the plane of the figure. The RF angle of arrival is denoted by $\theta$, while $\mathrm{\beta=\angle(\mathbf{E}_{RF},\mathbf{E}_{opt})}$ denotes the angle between the RF and optical polarizations. For the geometry considered here, the AoA $\theta=90^\circ-\beta$. The transmitted probe beam is detected by a photodetector to obtain the EIT spectrum. (b) Rydberg-EIT excitation scheme for the $6S_{1/2} \rightarrow 6P_{3/2} \rightarrow 42D_{5/2}$ ladder. The incident off-resonant RF field produces $m_J$-dependent AC Stark shifts, resulting in  spectrally resolvable $|m_J|$ branches.}
    \label{fig:Levels_and_Exp_Setup}
\end{figure}

Rydberg-atom sensors promise to provide broadband reception capabilities with electrically small devices \cite{holloway_broadband_2014,sedlacek_atom-based_2013,fan_subwavelength_2014}. Taking advantage of this fact, it is possible to measure the amplitude and phase of RF fields within sub-wavelength apertures \cite{jing_atomic_2020, simons_rydberg_2019}. AoA determination using RF phase measurements at two spatial locations within a Rydberg vapor cell  is demonstrated  in Ref \cite{robinson_determining_2021}, while a technique to determine the AoA using only two amplitude measurements of a standing wave formed above a metallic plate in a Rydberg vapor cell is shown in \cite{talashila_determining_2025}. A different way to determine incident angle is to use fluorescence imaging of a 2D standing wave in a glass vapor cell with Rydberg atoms \cite{schlossberger_angle--arrival_2025}. A recent study demonstrated simultaneous AoA measurements of multiple sources \cite{gill_microwave_2026}.



Under suitable laser excitation, Rydberg atoms can exhibit electromagnetically induced transparency (EIT), a quantum interference phenomenon in which the absorption of a probe laser is suppressed by the presence of a coupling laser, resulting in enhanced probe transmission through the atomic medium. When the applied RF field is resonant with a transition between Rydberg states, the EIT spectrum exhibits Autler–Townes (AT) splitting \cite{autler_stark_1955}, with the splitting proportional to the amplitude of the applied electric field. In contrast, for off-resonant RF fields, the Rydberg levels undergo AC Stark shifts, which, in the perturbative regime, are proportional to the square of the applied electric-field amplitude. The degenerate ($m_J$) sublevels split in the presence of an applied RF field due to their different polarizabilities \cite{autler_stark_1955,zimmerman_stark_1979,miller_radio-frequency-modulated_2016,anderson_optical_2016}. With the aid of theoretically calculated polarizability values, the AC Stark shifts of the respective peaks can be used to determine a calibrated incident RF field amplitude \cite{song_continuous_2024}. Previous works have demonstrated the use of dc Stark-shifted $m_J$ amplitudes for determining the orientation of static dc electric fields \cite{li_super_2023, behary_effects_2026}. In this work, we show that the relative spectral amplitudes of the AC Stark-shifted $m_J$ states depend on the angle between the polarization axes of the incident RF field and the Rydberg excitation lasers, and that this can be used to determine the angle of arrival (AoA) of the incident RF wave. This approach  provides a means of determining the AoA from a single spatially localized EIT spectrum under the controlled polarization geometry considered here without the need for phase measurements of the RF field.

\section{Experimental Setup}

\begin{figure}
    \centering
    \includegraphics[width=0.99\linewidth]{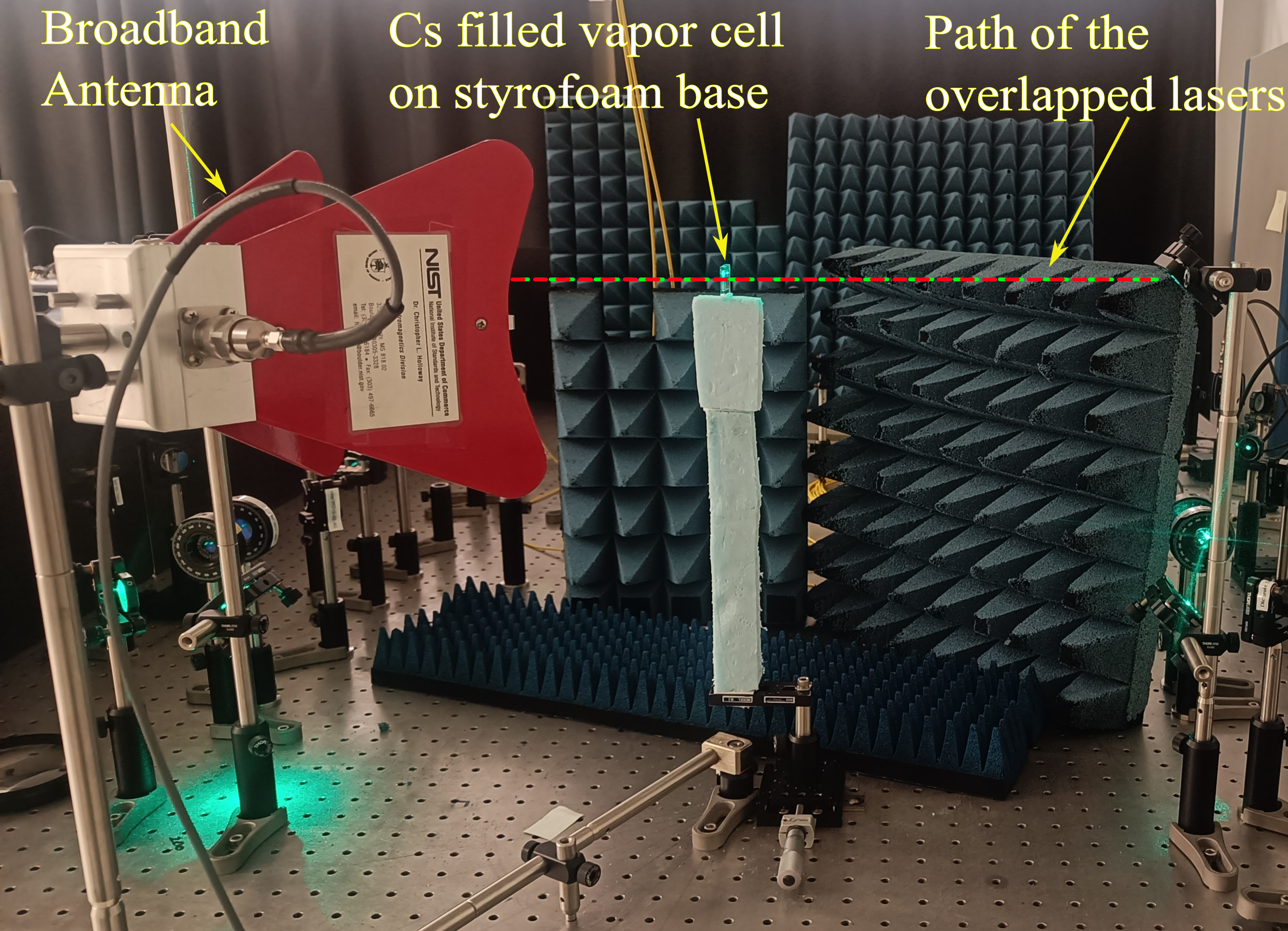}
    \caption{Photograph of the experimental setup. The RF horn antenna is mounted on a rotating support beneath the vapor cell for varying the AoA $\theta$ while maintaining a fixed antenna-to-cell distance. The vapor cell is supported on a styrofoam base, and the counter-propagating probe and coupling beams are spatially overlapped within the cell.}
    \label{fig:Photo_Experiment}
\end{figure}

The experimental investigation uses a room-temperature cesium vapor cell to study the RF-induced Stark shift of the $42D_{5/2}$ Rydberg state via EIT. The vapor cell has internal dimensions of $20 \times 15 \times 2\,\mathrm{mm}^3$ and contains cesium vapor enclosed within $1\,\mathrm{mm}$-thick glass  walls \cite{artusio-glimpse_batch-fabrication_2026}. The schematic of the experimental setup is shown in Fig.\,\ref{fig:Levels_and_Exp_Setup}(a), while a photograph of the actual setup is presented in Fig.\,\ref{fig:Photo_Experiment}. In the experiment, the vapor cell is placed on a styrofoam support and surrounded, to the extent possible, by RF absorbers to minimize reflections and scattering; residual RF-field inhomogeneity and reflections are not explicitly included in the model. A two-photon ladder-type EIT scheme is implemented using counter-propagating probe and coupling laser beams as shown in Fig.\,\ref{fig:Levels_and_Exp_Setup}(b). The probe laser is frequency-locked to the cesium $6S_{1/2}(F=4)\rightarrow6P_{3/2}(F'=5)$ transition at $852\,\mathrm{nm}$ and delivers approximately $15\,\mu\mathrm{W}$ of optical power with a $1/e^2$ beam diameter of $350\,\mu\mathrm{m}$ corresponding to a Rabi rate of $\Omega_p/2\pi = 8.02 \,\mathrm{MHz}$. The coupling laser at $509\,\mathrm{nm}$ is tuned to the $6P_{3/2}(F'=5)\rightarrow42D_{5/2}$ transition and delivers approximately $22\,\mathrm{mW}$ of optical power with a  $1/e^2$ beam diameter of $450\,\mu\mathrm{m}$ corresponding to a Rabi rate of $\Omega_c/2\pi = 1.56 \,\mathrm{MHz}$. Both lasers are linearly polarized and spatially overlapped within the vapor cell. An RF horn antenna is oriented such that its electric field is polarized in the same plane as the laser light polarizations at all angles of incidence. As shown in Fig.~\ref{fig:Levels_and_Exp_Setup}(a), the electric fields within the vapor cell lie in the plane of this paper.

The transmitted probe beam through the vapor cell is detected using a photodetector and recorded with a digital acquisition system while the coupling laser frequency is scanned across the Rydberg resonance. Typical scan ranges of $\pm100\,\mathrm{MHz}$ around resonance are used to resolve the AC Stark-shifted EIT spectral features. Measurements are acquired from $\theta = 0^{\circ}$ to $70^{\circ}$ in  $10^{\circ}$ increments  while maintaining constant applied RF power, antenna-to-cell distance, and optical alignment. For $\theta>70^{\circ}$, the horn antenna obstructs the optical beam paths, preventing the measurements at larger angles. For each angular configuration, the coupling laser frequency is scanned and the transmitted probe signal is averaged over 10 acquisitions to improve the signal-to-noise ratio. The measured EIT spectra exhibited multiple RF-dressed resonances corresponding to the Stark-shifted $\left|m_J\right|=1/2$, $3/2$, and $5/2$ sublevels of the $42D_{5/2}$ state as shown in Fig. \ref{fig:EIT_Spectrums}(a). In the geometry shown in Fig.\,\ref{fig:Levels_and_Exp_Setup}(a), the incident RF field is linearly polarized along $\mathrm{\mathbf{E}_{RF}}$, while the probe and coupling beams are linearly co-polarized and define a common optical polarization axis $\mathrm{\mathbf{E}_{opt}}$. Rotation of the antenna about the cell changes the angle $\beta=\angle(\mathrm{\mathbf{E}_{RF}},\mathrm{\mathbf{E}_{opt}})$ while maintaining the optical geometry. For the configuration used here, the relation $\theta=90^\circ-\beta$, provides a one-to-one mapping between the polarization angle and the magnitude of the AoA.



The $m_J$-dependent Stark shift of the EIT spectrum is given by \cite{bai_precise_2020,sibalic_arc_2017}

\begin{equation}
    \Delta \nu_{AC} (m_J,E) = - \cfrac{E^2}{2h}
                                \left[ 
                                \alpha_0 + \alpha_2 \cfrac{3m_J^2-J(J+1)}{J(2J-1)} 
                                \right],
\end{equation}
where $\alpha_0$ and $\alpha_2$ correspond to the frequency-dependent scalar and tensor polarizabilities, respectively, and $h$ is the Planck constant. Here, $\Delta \nu_{AC}$ denotes the AC Stark shift, $J$ is the total electronic angular momentum quantum number , $m_J$ is its projection along the quantization axis, and $E$ is the rms RF electric field amplitude. The vector polarizability $\alpha_1$ does not contribute for the linearly polarized RF field used here. This equation provides a perturbative description of $m_J$-dependent Stark shifts. However, to describe angle-dependent spectral features, a simplified model with Floquet-Shirley 
calculations is presented below.

\section{Modeling}

\begin{figure*}
    \centering
    \includegraphics[width=0.51\linewidth]{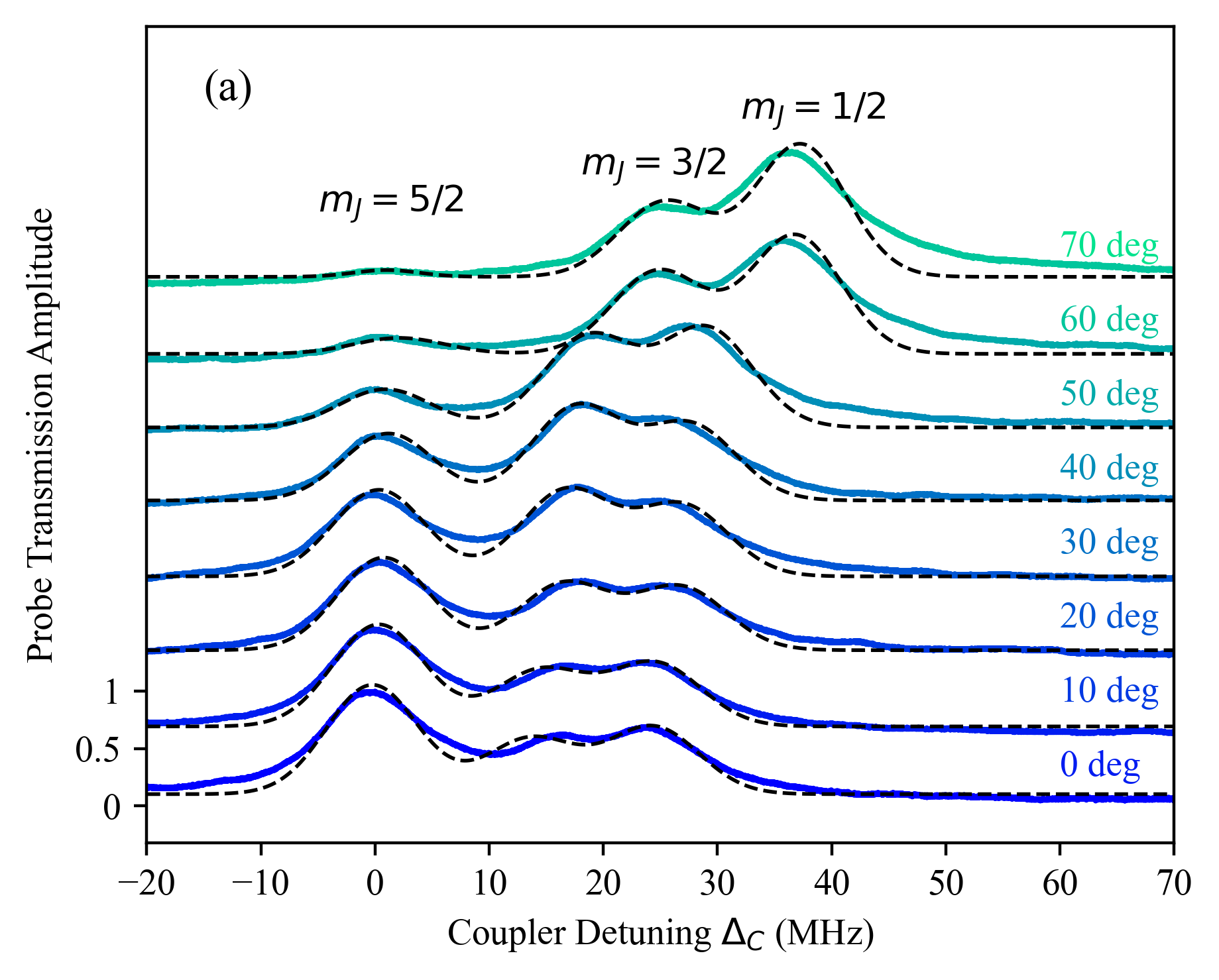}\hfill
    \includegraphics[width=0.48\linewidth]{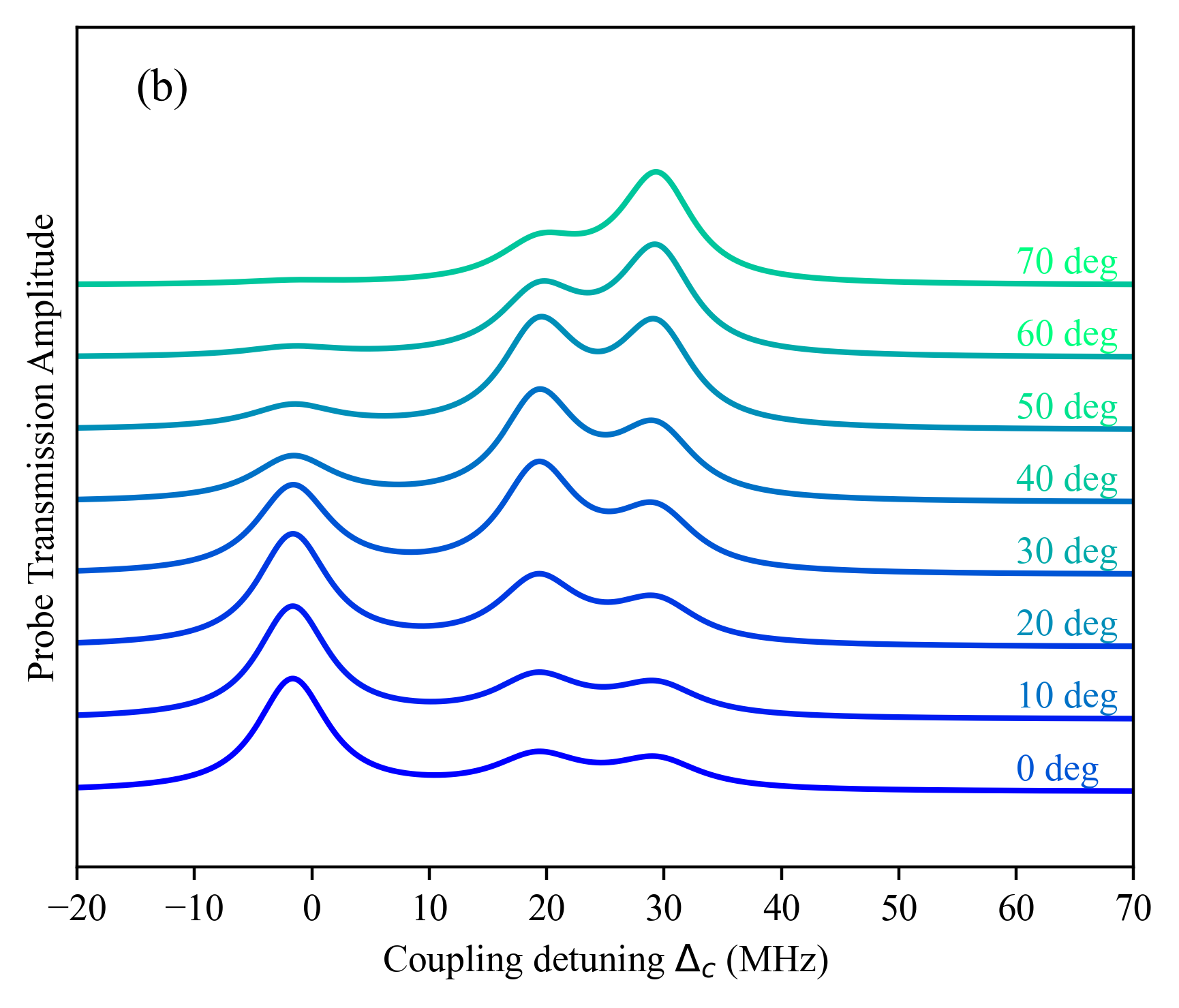}
    \caption{(a) Measured AC Stark-shifted EIT spectra as a function of coupling-laser detuning ($\Delta_c$), for various AoAs at an RF frequency of $1.27\,\mathrm{GHz}$. Solid curves show the measured spectra and dashed curves show Gaussian fits to the $|m_J|=1/2$, $3/2$, and $5/2$ resonances. The traces are vertically offset for visual clarity. (b) Calculated spectra for the same RF frequency and angular configurations. The model reproduces the principal angle-dependent redistribution of spectral amplitude among the $|m_J|$ branches. }
    \label{fig:EIT_Spectrums}
\end{figure*}

The measured EIT spectra as a function of the coupling-laser detuning, $\Delta_c$, for various RF incident angles are shown in Fig.\,\ref{fig:EIT_Spectrums}(a). The plots are vertically offset for visual clarity. The dashed curves represent Gaussian fits to the three $m_J$ peaks. The fitted peak positions and amplitudes are used to extract the AC Stark shifts and  to compare the measured spectra with Floquet-Shirley simulations to investigate the influence of the optical/RF geometry on the observed magnetic sublevel spectral amplitudes. The numerical model describes a two-photon ladder excitation scheme in cesium involving the transitions $6S_{1/2}(F=4)\rightarrow 6P_{3/2}(F'=5)\rightarrow 42D_{5/2}$ in the presence of an RF electric field. The RF-induced AC Stark shifts of the Rydberg sublevels are calculated using the Floquet-Shirley formalism implemented in the ARC (\textit{Alkali Rydberg Calculator}) package \cite{sibalic_arc_2017}. The quantization axis is chosen along the RF electric-field polarization. For each $\left|m_J\right|=1/2,\,3/2,$ and $5/2$ branch of the $42D_{5/2}$ manifold, a dressed-state basis is constructed including nearby dipole-coupled Rydberg states over a finite basis with $(n=42\pm6,\,l<6)$. 
With the quantization axis chosen along the RF polarization, the calculated RF-dressed energies are essentially independent of the AoA ($\theta$), whereas the optical excitation amplitudes depend on this angle.



The angle-dependent excitation strengths of the individual branches $\left|m_J\right|$ are calculated using a coherent hyperfine-to-fine-structure angular momentum model. In this approach, the intermediate hyperfine state $6P_{3/2}(F'=5,m_F)$ is expanded in the coupled basis $\left|J_e,m_J;I,m_I\right\rangle$, and the two-photon excitation amplitudes are obtained by coherently summing all allowed  optical pathways, weighted by the appropriate Clebsch--Gordan coefficients and polarization projections \cite{fan_atom_2015}. This treatment captures the experimentally observed redistribution of spectral weight among the $\left|m_J\right|=1/2,\,3/2,$ and $5/2$ branches as the AoA varies from $0^\circ$ to $70^\circ$. 

The combined model shown in Fig.\,\ref{fig:EIT_Spectrums}(b) captures the  overall redistribution of spectral amplitudes among the magnetic sublevels arising from the changing projection of the optical polarization onto the RF-defined quantization axis.  The qualitative agreement confirms that the dominant mechanism governing the angular dependence is the coherent coupling of the optical fields to the RF-dressed magnetic sublevels. 

\section{Results and Discussion}
\begin{figure*}
    \centering
    \includegraphics[width=0.99\linewidth]{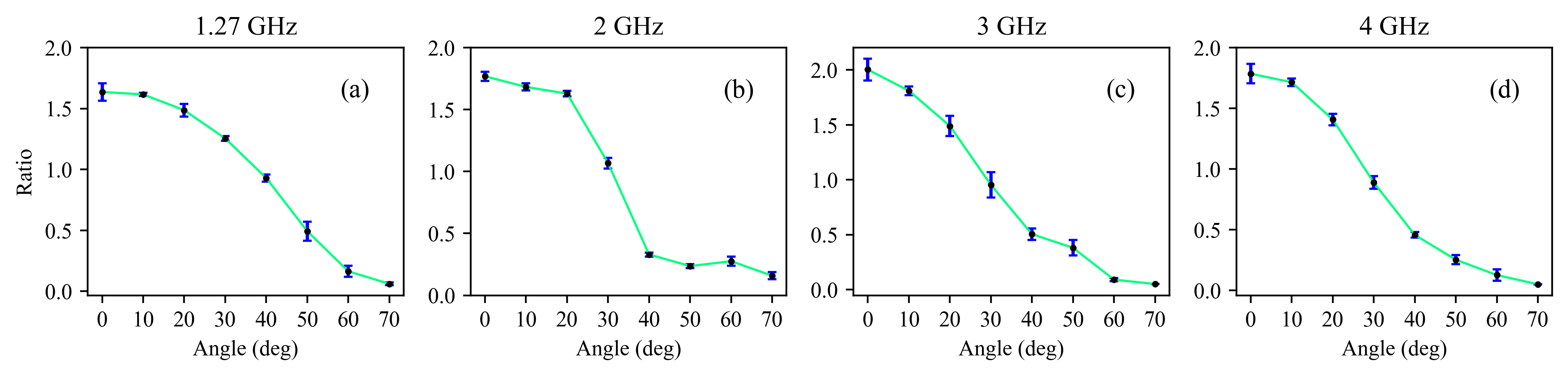}
    \caption{Experimentally obtained ratio $R(\theta) = A_{|m_J|=5/2}/A_{|m_J|=1/2}$ at RF frequencies of (a) 1.27 GHz, (b) 2 GHz, (c) 3 GHz and (d) $4\,\mathrm{GHz}$. Each point represents the mean of five independent measurements; error bars indicate $\pm 1$ standard deviation.}
    \label{fig:AoA_Ratios}
\end{figure*}

Peak positions of the EIT spectra shown in Fig. \ref{fig:EIT_Spectrums}(a) are extracted by fitting Gaussian functions to the individual resonances. The measured RF-dressed EIT spectra exhibit three distinct Stark-shifted branches corresponding to the $|m_J|=1/2$, $3/2$, and $5/2$ sublevels of the $42D_{5/2}$ Rydberg state. The angular dependence of the measured ratio $R(\theta) = A_{|m_J|=5/2}/A_{|m_J|=1/2}$ between the $|m_J|=5/2$ and $|m_J|=1/2$ spectral amplitudes is used to infer the AoA. As the angle between the optical polarization and the RF electric-field direction is varied, the relative amplitudes of the three $m_J$ peaks change. In particular, the $|m_J|=5/2$ amplitude dominates for small $\theta$, for which the incident RF field polarization is nearly orthogonal to the laser polarizations. However, the $|m_J|=1/2$ peak becomes dominant as the incident RF field polarization approaches parallel with the laser polarizations. The relative amplitudes of the individual $|m_J|$ peaks in the AC Stark-shifted EIT spectrum vary strongly with $\theta$,  and their fitted resonance positions exhibit only weak angular dependence. This behavior is consistent with the expectation that the RF field  determines the dressed-state energies, whereas the optical polarization primarily controls the excitation probability of the individual magnetic sublevels.

Although the model reproduces the major experimental trends, quantitative differences remain. Several physical effects may contribute to these differences. First, the simulation neglects optical pumping and population redistribution among the ground-state hyperfine and Zeeman sublevels, which can significantly modify the steady-state population available for excitation. Second, the model assumes a homogeneous RF field and uniform optical intensity, whereas spatial averaging over the interaction region can alter the relative visibility of individual dressed-state resonances. Third, residual magnetic fields, imperfect polarization purity, and small misalignment between the optical and RF fields can mix magnetic sublevels and modify the effective excitation strengths. 

The experimental resonances are broader than the linewidths in the simplified calculated spectra. The possible contributions to this broadening include laser phase noise, transit-time effects, collisional dephasing, residual Doppler averaging, and RF-field inhomogeneity. These effects are not explicitly included in the present calculation and therefore primarily influence the measured resonance widths rather than the resonance positions. Nevertheless, the calculated spectra provide a useful framework for interpreting the angular dependence of the RF-dressed EIT signal and for identifying the underlying magnetic sublevel structure of the observed resonances.

The angular dependence of the  measured ratio $R(\theta) = A_{5/2}/A_{1/2}$  is shown in Fig.~\ref{fig:AoA_Ratios} for RF frequencies of $1.27$, $2$, $3$, and $4\,\mathrm{GHz}$. Each spectrum is obtained by averaging ten frequency scans, and the amplitude ratios in Fig.\ref{fig:AoA_Ratios} are subsequently averaged over five independently acquired spectra. For all investigated frequencies, $R(\theta)$ exhibits an overall decreasing trend with increasing AoA; small deviations from strict monotonicity are consistent with the experimental uncertainty. At small angles ($0^\circ$--$20^\circ$), the $\left|m_J\right|=5/2$ resonance is significantly stronger than the $\left|m_J\right|=1/2$ resonance, yielding ratios between approximately $1.5$ and $2.0$ depending on the RF frequency. As the angle increases, the relative strength of the $\left|m_J\right|=5/2$ feature progressively decreases, and the ratio approaches unity near $30^\circ$--$40^\circ$. At larger angles, the $|m_J|=1/2$ branch becomes dominant, causing the ratio to fall below 0.5 and eventually approach zero near $70^\circ$. 

The local $1\sigma$ AoA uncertainty is estimated as $\sigma_\theta = \sigma_R/|dR_{fit}/d\theta|$, where $\sigma_R$ is the measured standard deviation of $R(\theta)$ and $R_{fit}(\theta)$ is a smooth monotonic cubic-spline curve. Over the $20^\circ - 40^\circ$ range, the mean estimated local $1\sigma$ uncertainties are approximately $1.5^\circ$, $0.6^\circ$, $2.0^\circ$ and $1.0^\circ$ at 1.27, 2, 3, and $\mathrm{4\,GHz}$, respectively. These values are uncertainty estimates and are not directly measured angular resolutions as the measurements were performed at $10^\circ$ angular intervals. These estimates characterize measurement repeatability and do not include systematic uncertainties associated with antenna positioning, polarization alignment, or the fitted calibration curves.

Although the overall angular trend is consistent across the entire frequency range, subtle differences in the magnitude and slope of the ratio are evident. 


The transition from $\left|m_J\right|=5/2$-dominated to $\left|m_J\right|=1/2$-dominated spectra occurs over a similar angular range for all frequencies, suggesting that the geometrical projection of the optical polarization onto the  quantization axis along the RF linear polarization is the dominant mechanism governing the observed behavior. The decrease of $R(\theta)$ toward zero at large angles indicates preferential excitation of the $|m_J|=1/2$ branch as the optical and RF polarizations approach parallel alignment. The consistency of this overall angular dependence over the RF frequency range $1.27 -4\,\mathrm{GHz}$ demonstrates the robustness of the angular response and supports its potential application for angle-of-arrival estimation and vector electric-field sensing using RF-dressed Rydberg-atom spectroscopy. For a known RF carrier frequency, the ratio $R(\theta)$ provides an estimate of AoA. However, as it does not distinguish the angles $+\theta$ and $-\theta$, this method provides an estimate only of the magnitude of the AoA over the range $0^\circ$ to $70^\circ$.


\section{Conclusion}
We experimentally investigated RF-induced Stark shifts of the cesium $42D_{5/2}$ Rydberg state using a two-photon ladder EIT scheme in the presence of an RF field in the frequency range of  $1.27$ to $4$ GHz. By varying the angle between the optical polarization and the RF field, we observed substantial changes in the relative amplitudes of the $\left|m_J\right|=1/2$, $3/2$, and $5/2$ spectral features while the AC Stark-shifted resonance positions remained nearly unchanged. The results demonstrate that the RF interaction determines the dressed-state energies, whereas the relative orientation of the RF and optical polarizations governs the excitation strengths of the individual magnetic sublevels. To interpret the measurements, we developed a numerical model combining Floquet-Shirley calculations with a coherent hyperfine-to-fine-structure excitation framework is developed. The model qualitatively reproduces the experimentally observed AC Stark shifts and captures the principal angular dependence of the spectra. Minor discrepancies indicate that additional physical mechanisms, including optical pumping, field inhomogeneity, polarization imperfections, and higher-order state mixing, may play an important role in determining the detailed amplitude distribution among the dressed states. The present demonstration determines the magnitude of the AoA within a calibrated one-dimensional angular sector under a known linear-polarization geometry.

We demonstrate a Rydberg AoA-determination method based on AC Stark shifts in a compact $20 \times 15 \times 2\,\mathrm{mm}^3$  subwavelength vapor cell. This method requires only an AC Stark-shifted EIT spectrum measured at a single spatial location and does not require measurement of the incident RF phase. We experimentally demonstrate the method with a cesium-filled vapor cell over the RF frequency range $1.27-4\,\mathrm{GHz}$. For future work, we aim to extend the experimental realization to a broader frequency range, study the effects of the principal quantum number of the Rydberg excitation on the measurement results, and examine the effects of other experimental parameters. The approach provides a path toward compact AoA sensing over a broad frequency range based on a single spatially localized Rydberg-atom measurement.


\section*{data availability}
   All of the data presented in this paper and used to
support the conclusions of this article is available at \cite{holloway_data_nodate}. 

\begin{acknowledgments}
A contribution of the U.S. government, this work is not subject to copyright in the U.S.
\end{acknowledgments}

\bibliography{Refs}

\end{document}